# Modulation instabilities in birefringent two-core optical fibers

**J. H. Li**[1], **K. S. Chiang**[2*], **B. A. Malomed**[3†], and **K. W. Chow**[1‡]

[1]*Department of Mechanical Engineering, University of Hong Kong, Hong Kong*

[2]*Department of Electronic Engineering, City University of Hong Kong, Hong Kong*

[3]*Department of Physical Electronics, School of Electrical Engineering,*

*Tel Aviv University, Tel Aviv 69978, Israel*

[*] Email: eeksc@cityu.edu.hk

[†] Email: malomed@post.tau.ac.il

[‡] **Corresponding author** - FAX: (852) 2858 5415; Email: kwchow@hku.hk

**Abstract**

Previous studies of the modulation instability (MI) of continuous waves (CWs) in a two-core fiber (TCF) did not consider effects caused by co-propagation of the two polarized modes in a TCF that possesses birefringence, such as cross-phase modulation (XPM), polarization-mode dispersion (PMD), and polarization-dependent coupling (PDC) between the cores. This paper reports an analysis of these effects on the MI by considering a linear-birefringence TCF and a circular-birefringence TCF, which feature different XPM coefficients. The analysis focuses on the MI of the asymmetric CW states in the TCFs, which have no counterparts in single-core fibers. We find that, the asymmetric CW state exists when its total power exceeds a threshold (minimum) value, which is sensitive to the value of the XPM coefficient. We consider, in particular, a class of asymmetric CW states that admit analytical solutions. In the anomalous dispersion regime, without taking the PMD and PDC into account, the MI gain spectra of the birefringent TCF, if scaled by the threshold power, are almost identical to those of the zero-birefringence TCF. However, in the normal dispersion regime, the power-scaled MI gain spectra of the birefringent TCFs are distinctly different from their zero-birefringence counterparts, and the difference is particularly significant for the circular-birefringence TCF, which takes a larger XPM coefficient. On the other hand, the PMD and PDC only exert weak effects on the MI gain spectra. We also simulate the nonlinear evolution of the MI of the CW inputs in the TCFs and obtain a good agreement with the analytical solutions.

# 1. Introduction

Modulation instability (MI) of plane or continuous waves (CWs) arises in many fields of physics, e.g. Bose-Einstein condensates [1, 2], fluid mechanics [3] and optics [4, 5]. This study focuses on optical fibers, where the dynamics is determined by the interplay of dispersive and nonlinear effects, as has been demonstrated in many diverse settings [4–9]. The MI is a physically important problem, which is closely related to the Fermi-Pasta-Ulam recurrence effect and the formation of solitons [5, 10].

The MI in a conventional single-core fiber has been studied thoroughly [4–9]. If the wave propagation in the fiber is governed by the nonlinear Schrödinger (NLS) equation, which takes into account the self-phase modulation (SPM) and the group-velocity dispersion (GVD), the MI occurs only in the anomalous dispersion regime [5]. However, the MI can also be induced in the normal dispersion regime by other effects, such as cross-phase modulation (XPM) in the case of bimodal CWs [11], higher even-order dispersion [12], dispersive losses (alias spectral filtering) [13], and birefringence [14–18]. The effects of linear birefringence have drawn special attention, as linear birefringence can also generate new MI bands in the anomalous dispersion regime.

This paper addresses the MI in a two-core fiber (TCF), i.e., a fiber that consists of two linearly coupled identical parallel cores. The phenomenon of periodic optical power transfer between the two cores along a TCF [19] is widely used in many practical fiber-optic devices. Various aspects of the MI in a TCF have been studied [20–22]. The MI characteristics of the symmetric and antisymmetric CW states in a TCF are qualitatively similar to those in a conventional single-core fiber [20]. On the other hand, the spontaneous symmetry breaking in linearly coupled systems gives rise to asymmetric CW states in a TCF [23], which makes its MI characteristics qualitatively different from those of a single-core fiber [21]. In particular, the dispersion (or wavelength dependence) of the coupling coefficient between the two cores can drastically modify the MI bands of the asymmetric states in both the anomalous and normal dispersion regimes [22].

All the previous studies of the MI in a TCF [20–22] ignored potentially significant effects caused by co-propagation of the two polarized modes in a TCF that possesses birefringence, such as XPM, polarization-mode dispersion (PMD), and polarization-dependent coupling (PDC) between the cores. In reality, TCFs, especially the recent ones based on photonic-crystal structures [24], can exhibit strong linear birefringence. The objective of the present work is to study the effects of birefringence on the MI characteristics of a TCF, by considering a linear-birefringence TCF and a circular-birefringence TCF. A linear-birefringence TCF, where each core supports two orthogonal linearly polarized modes, is just the ordinary TCF used nowadays (see, for example, [24]). A circular-birefringence TCF, where each core supports the right- and left-circularly polarized modes, is a special fiber that could be fabricated by making the two cores rotate around the central axis of the fiber rapidly (e.g., by rapidly spinning the fiber preform during the fiber drawing process or by strongly twisting an ordinary TCF) [25]. The circular-birefringence TCF features much stronger XPM than the linear-birefringence TCF and, therefore, a comparison of their MI characteristics can help to highlight the significance of the XPM effects.

In this paper, the asymmetric CW states are considered, rather than the symmetric/antisymmetric CW states, for which the situation is too similar to previously studied cases [20]. We find that, the asymmetric CW state emerges when the total input power exceeds a minimum (threshold) value, which strongly depends on the XPM coefficient. While the most general asymmetric CW states are not tractable analytically, we consider a special class of asymmetric CW states that admit analytical solutions. In the anomalous dispersion regime, without considering the PMD and PDC effects, the MI gain spectra of both birefringent TCFs are almost identical to those of the zero-birefringence TCF, if scaled by the respective threshold powers. However, in the normal dispersion regime, the power-scaled MI gain spectra of the birefringent TCFs are notably different from those of the zero-birefringence TCF and the difference is much stronger for the circular-birefringence TCF. On the other hand, the PMD and PDC in these fibers only have a weak influence on the MI characteristics. In addition, we verify the predictions on the dominant unstable mode from the analytical MI solutions, by direct numerical simulations of the coupled NLS equations.

**2. Coupled-mode equations and the analysis of modulation instability**

*2.1 Coupled-mode equations*

In the high-birefringence limit for the TCF, the propagation of slowly varying amplitudes of the electric fields along coordinate *z* is described by four coupled generalized NLS equations:

$$i\left(\frac{\partial a_{1x}}{\partial z} + \beta_{1x}\frac{\partial a_{1x}}{\partial t}\right) - \frac{1}{2}\beta_{2x}\frac{\partial^2 a_{1x}}{\partial t^2} + \gamma_x\left(|a_{1x}|^2 + \sigma|a_{1y}|^2\right)a_{1x} + C_x a_{2x} + iC_{1x}\frac{\partial a_{2x}}{\partial t} = 0,$$

$$i\left(\frac{\partial a_{2x}}{\partial z} + \beta_{1x}\frac{\partial a_{2x}}{\partial t}\right) - \frac{1}{2}\beta_{2x}\frac{\partial^2 a_{2x}}{\partial t^2} + \gamma_x\left(|a_{2x}|^2 + \sigma|a_{2y}|^2\right)a_{2x} + C_x a_{1x} + iC_{1x}\frac{\partial a_{1x}}{\partial t} = 0,$$

$$i\left(\frac{\partial a_{1y}}{\partial z} + \beta_{1y}\frac{\partial a_{1y}}{\partial t}\right) - \frac{1}{2}\beta_{2y}\frac{\partial^2 a_{1y}}{\partial t^2} + \gamma_y\left(|a_{1y}|^2 + \sigma|a_{1x}|^2\right)a_{1y} + C_y a_{2y} + iC_{1y}\frac{\partial a_{2y}}{\partial t} = 0,$$

$$i\left(\frac{\partial a_{2y}}{\partial z} + \beta_{1y}\frac{\partial a_{2y}}{\partial t}\right) - \frac{1}{2}\beta_{2y}\frac{\partial^2 a_{2y}}{\partial t^2} + \gamma_y\left(|a_{2y}|^2 + \sigma|a_{2x}|^2\right)a_{2y} + C_y a_{1y} + iC_{1y}\frac{\partial a_{1y}}{\partial t} = 0. \quad (1)$$

Here $a_{mj}$ ($m = 1, 2$ and $j = x, y$) are the amplitudes of the $j$ polarization in the $m$-th core, $\beta_{1j}$ is the group delay of the respective polarization, $\beta_{2j}$ is its GVD coefficient at the carrier frequency ($\beta_{2j}$ are negative and positive for anomalous and normal dispersion, respectively), $\gamma_j$ is the nonlinearity coefficient of the $j$ polarization, $\sigma$ is the relative XPM coefficient, $C_j$ is the linear coupling coefficient responsible for the power exchange between the two cores in the $j$ polarization, and $C_{1j} = dC_j/d\omega$, evaluated at the carrier frequency $\omega$, represents the dispersion of the coupling coefficient in the $j$ polarization. The latter effect is equivalent to the intermodal dispersion arising from the group-delay difference between the even and odd supermodes of the TCF [26, 27]. Nonlinear coupling between the two cores is ignored in Eq. (1), which is justified by the fact that, in most practical situations, the spatial overlap between the fields of the modes propagating in the two cores is negligibly small. Nonlinear coupling may need to be included, however, in unusual situations where the two cores are very close to each other and the modes are operated in the close-to-cutoff regime (i.e., when the two cores are exceptionally strongly coupled in the linear sense).

Subscripts *x* and *y*, attached to the polarization components in Eq. (1), naturally refer to the orthogonal linearly polarized (in the *x*- and *y*-directions) modes of the individual cores of the linear-birefringence TCF, whose XPM coefficient is $\sigma = 2/3$ [28]. For the sake of the uniformity of the notation, we apply the same subscripts, *x* and *y*, to the clockwise and counter-clockwise circularly polarized modes of the individual cores of the circular-birefringence TCF, whose XPM coefficient is $\sigma = 2$ [28]. As usual, rapidly oscillating four-wave mixing terms are neglected in Eq. (1) [5].

Usually, the polarization dependences of the GVD, the nonlinearity coefficient, and the coupling coefficient dispersion, are weak. Consequently, we set $\beta_{2x} = \beta_{2y} \equiv \beta_2$, $\gamma_x = \gamma_y \equiv \gamma$, and $C_{1x} = C_{1y} \equiv C_1$. The PMD and PDC coefficients in the TCF are defined as $\Gamma \equiv \beta_{1x} - \beta_{1y}$ and $\Delta C \equiv C_x - C_y$. The objective of the study is to understand how the polarization-dependent parameters $\sigma$, $\Gamma$, and $\Delta C$ affect the MI characteristics of the TCF, when CWs carried by both polarized modes of the fiber are launched into the fiber.

It is relevant to mention that a system of four coupled equations meant for a birefringent TCF was postulated in some earlier works in the literature [29], but the system there does not include any linear coupling between the two cores, and hence cannot be considered as a valid model for a TCF. As shown in the following sections (and is obvious anyway), the linear coupling coefficients, namely $C_x$ and $C_y$, which are absent in the earlier models [29], play a central role in the analysis of the MI characteristics of a birefringent TCF.

*2.2 Modulation instability analysis*

CW solutions to Eq. (1) depend on two different propagation constants, $k_1$ and $k_2$, which pertain to the different polarization components, while a given polarization must be, obviously, carried by the same propagation constant in both cores:

$$a_{1x} = A_1 \exp(ik_1z), \quad a_{2x} = A_2 \exp(ik_1z), \quad a_{1y} = B_1 \exp(ik_2z), \quad a_{2y} = B_2 \exp(ik_2z). \quad (2)$$

Equations (1) and (2) admit one CW solution with the amplitudes obeying the following relations:

$$A_1 A_2 = \frac{C_x - \sigma C_y}{\gamma(1-\sigma^2)}, \quad B_1 B_2 = \frac{C_y - \sigma C_x}{\gamma(1-\sigma^2)}, \quad \frac{B_1}{A_1} = \frac{B_2}{A_2} = \sqrt{\frac{\sigma C_x - C_y}{\sigma C_y - C_x}}, \quad (3a)$$

and

$$k_1 = \frac{\gamma(1+\sigma)C_x}{C_x + C_y} P, \quad k_2 = \frac{\gamma(1+\sigma)C_y}{C_x + C_y} P, \quad (3b)$$

where $P = A_1^2 + B_1^2 + A_2^2 + B_2^2$ is the total CW power. The CW state given by Eqs. (2) and (3) is symmetric or antisymmetric when $A_1 = A_2$ or $A_1 = -A_2$, being asymmetric otherwise. Note that Eq. (3a) yields physically relevant (real) solutions under the conditions $\sigma C_x < C_y < C_x/\sigma$ for $0 \leq \sigma < 1$, and $\sigma C_x > C_y > C_x/\sigma$ for $\sigma \geq 1$.

In the CW solutions given by Eqs. (2) with (3), one of the amplitudes, for instance, $A_1$, may be chosen arbitrarily, as only three relations out of four in Eq. (3a) are independent. The most general CW asymmetric solutions of Eq. (1) depend on two independent parameters, namely the two propagation constants, $k_1$ and $k_2$, in Eq. (2). It is, however, impossible to find them analytically and hence study their MI in an analytical form. Therefore, we focus on the special analytical solutions given by Eqs. (2) and (3), which, as elaborated in Section III, are sufficient to demonstrate MI gain spectra that are notably different from their counterparts in a TCF without birefringence.

We also note that Eqs. (2) and (3) produce asymmetric CW solutions provided that the total CW power, $P$, exceeds a minimum (threshold) value, $P_{\min}(\sigma)$. With regard to Eq. (3a), this condition is cast into the following form:

$$P \equiv A_1^2 + B_1^2 + A_2^2 + B_2^2 = (1-\sigma)\frac{C_x + C_y}{C_x - \sigma C_y}\left(A_1^2 + A_2^2\right) \geq P_{\min}(\sigma) \equiv 2\frac{C_x + C_y}{\gamma(1+\sigma)}. \quad (4)$$

At the threshold, $P = P_{\min}(\sigma)$, one has $A_1 = A_2 = \sqrt{\frac{C_x - \sigma C_y}{\gamma(1-\sigma^2)}}$. The minimum power $P_{\min}(\sigma)$ is sensitive to the value of the XPM coefficient $\sigma$, the linear-birefringence TCF ($\sigma = 2/3$) has a higher threshold power than the circular-birefringence TCF ($\sigma = 2$). As shown in Section III, the minimum power plays a crucial role in the study of the MI gain spectra of the TCFs.

In the experiment, the transition to the asymmetric state can also be controlled, in an obvious way, by means of the total power of the beam coupled into the TCF. If the power exceeds $P_{\min}(\sigma)$, the asymmetric configuration will form by itself, through the instability of the symmetric state.

For a given total power, $P > P_{\min}(\sigma)$, the powers of the four components of the asymmetric CW solutions are

$$A_1^2 = \frac{C_x - \sigma C_y}{2(1-\sigma)(C_x + C_y)}\left(P \pm \sqrt{P^2 - P_{\min}^2(\sigma)}\right), \quad (5)$$

$$B_1^2 = \frac{C_y - \sigma C_x}{2(1-\sigma)(C_x + C_y)}\left(P \pm \sqrt{P^2 - P_{\min}^2(\sigma)}\right), \quad (6)$$

$$A_2^2 = \frac{C_x - \sigma C_y}{2(1-\sigma)(C_x + C_y)} \left( P \mp \sqrt{P^2 - P_{min}^2(\sigma)} \right), \tag{7}$$

$$B_2^2 = \frac{C_y - \sigma C_x}{2(1-\sigma)(C_x + C_y)} \left( P \mp \sqrt{P^2 - P_{min}^2(\sigma)} \right). \tag{8}$$

Hence, the power ratio between the two cores is

$$\frac{A_1^2 + B_1^2}{A_2^2 + B_2^2} = \frac{P \pm \sqrt{P^2 - P_{min}^2(\sigma)}}{P \mp \sqrt{P^2 - P_{min}^2(\sigma)}} = \frac{R \pm \sqrt{R^2 - 1}}{R \mp \sqrt{R^2 - 1}}, \tag{9}$$

where $R = P/P_{min}(\sigma)$ is the total power normalized to the minimum power.

To study the stability of the CW state, we seek perturbed solutions in the following form:

$$a_{1x} = (A_1 + u_1)\exp(ik_1 z), \quad a_{2x} = (A_2 + u_2)\exp(ik_1 z)$$
$$a_{1y} = (B_1 + v_1)\exp(ik_2 z), \quad a_{2y} = (B_2 + v_2)\exp(ik_2 z), \tag{10}$$

where $u_i \equiv u_i(z, t)$ and $v_i \equiv v_i(z, t)$ ($i = 1, 2$) are small perturbations. With Eq. (10) inserted into Eq. (1), the linearization with respect to $u_i$ and $v_i$ yields

$$i\left(\frac{\partial u_1}{\partial z} + \beta_{1x}\frac{\partial u_1}{\partial t}\right) - \frac{1}{2}\beta_2\frac{\partial^2 u_1}{\partial t^2} + iC_1\frac{\partial u_2}{\partial t} + C_x u_2$$
$$+ \gamma\left[(A_1^2 - A_2^2 - \sigma B_2^2)u_1 + A_1^2 u_1^* + \sigma A_1 B_1(v_1 + v_1^*)\right] = 0,$$

$$i\left(\frac{\partial u_2}{\partial z} + \beta_{1x}\frac{\partial u_2}{\partial t}\right) - \frac{1}{2}\beta_2\frac{\partial^2 u_2}{\partial t^2} + iC_1\frac{\partial u_1}{\partial t} + C_x u_1$$
$$+ \gamma\left[(A_2^2 - A_1^2 - \sigma B_1^2)u_2 + A_2^2 u_2^* + \sigma A_2 B_2(v_2 + v_2^*)\right] = 0,$$

$$i\left(\frac{\partial v_1}{\partial z} + \beta_{1y}\frac{\partial v_1}{\partial t}\right) - \frac{1}{2}\beta_2\frac{\partial^2 v_1}{\partial t^2} + iC_1\frac{\partial v_2}{\partial t} + C_y v_2$$
$$+ \gamma\left[(B_1^2 - B_2^2 - \sigma A_2^2)v_1 + B_1^2 v_1^* + \sigma A_1 B_1(u_1 + u_1^*)\right] = 0,$$

$$i\left(\frac{\partial v_2}{\partial z} + \beta_{1y}\frac{\partial v_2}{\partial t}\right) - \frac{1}{2}\beta_2\frac{\partial^2 v_2}{\partial t^2} + iC_1\frac{\partial v_1}{\partial t} + C_y v_1$$

$$+\gamma\left[(B_2^2 - B_1^2 - \sigma A_1^2)v_2 + B_2^2 v_2^* + \sigma A_2 B_2(u_2 + u_2^*)\right] = 0. \quad (11)$$

Further, we seek solutions for the perturbation modes in the following natural form:

$$u_1 = F_1 \exp(iKz - i\Omega t) + G_1 \exp(-iKz + i\Omega t),$$

$$u_2 = F_2 \exp(iKz - i\Omega t) + G_2 \exp(-iKz + i\Omega t),$$

$$v_1 = f_1 \exp(iKz - i\Omega t) + g_1 \exp(-iKz + i\Omega t),$$

$$v_2 = f_2 \exp(iKz - i\Omega t) + g_2 \exp(-iKz + i\Omega t), \quad (12)$$

where $F_i$, $G_i$, $f_i$, and $g_i$ ($i = 1, 2$) are real amplitudes, and $K$ and $\Omega$ are the wave number and the frequency of the perturbations. After substituting Eq. (12) into Eq. (11), the existence of nontrivial solutions for $F_i$, $G_i$, $f_i$ and $g_i$ requires the vanishing of the determinant of the corresponding coefficient matrix $M$, i.e.,

$$\det[M(K)] = 0 \quad (13)$$

with

$$M(K) =$$

$$\begin{pmatrix}
(A_1^2 - A_2^2 - \sigma B_2^2)\gamma + \beta_{1x}\Omega + \frac{1}{2}\beta_2\Omega^2 - K, & \gamma A_1^2, & C_x + C_1\Omega, & 0, & \gamma\sigma A_1 B_1, & \gamma\sigma A_1 B_1, & 0, & 0 \\
\gamma A_1^2, & (A_1^2 - A_2^2 - \sigma B_2^2)\gamma - \beta_{1x}\Omega + \frac{1}{2}\beta_2\Omega^2 + K, & 0, & C_x - C_1\Omega, & \gamma\sigma A_1 B_1, & \gamma\sigma A_1 B_1, & 0, & 0 \\
C_x + C_1\Omega, & 0, & (A_2^2 - A_1^2 - \sigma B_1^2)\gamma + \beta_{1x}\Omega + \frac{1}{2}\beta_2\Omega^2 - K, & \gamma A_2^2, & 0, & 0, & \gamma\sigma A_2 B_2, & \gamma\sigma A_2 B_2 \\
0, & C_x - C_1\Omega, & \gamma A_2^2, & (A_2^2 - A_1^2 - \sigma B_1^2)\gamma - \beta_{1x}\Omega + \frac{1}{2}\beta_2\Omega^2 + K, & 0, & 0, & \gamma\sigma A_2 B_2, & \gamma\sigma A_2 B_2 \\
\gamma\sigma A_1 B_1, & \gamma\sigma A_1 B_1, & 0, & 0, & (B_1^2 - B_2^2 - \sigma A_2^2)\gamma + \beta_{1y}\Omega + \frac{1}{2}\beta_2\Omega^2 - K, & \gamma B_1^2, & C_y + C_1\Omega, & 0 \\
\gamma\sigma A_1 B_1, & \gamma\sigma A_1 B_1, & 0, & 0, & \gamma B_1^2, & (B_1^2 - B_2^2 - \sigma A_2^2)\gamma - \beta_{1y}\Omega + \frac{1}{2}\beta_2\Omega^2 + K, & 0, & C_y - C_1\Omega \\
0, & 0, & \gamma\sigma A_2 B_2, & \gamma\sigma A_2 B_2, & C_y + C_1\Omega, & 0, & (B_2^2 - B_1^2 - \sigma A_1^2)\gamma + \beta_{1y}\Omega + \frac{1}{2}\beta_2\Omega^2 - K, & \gamma B_2^2 \\
0, & 0, & \gamma\sigma A_2 B_2, & \gamma\sigma A_2 B_2, & 0, & C_y - C_1\Omega, & \gamma B_2^2, & (B_2^2 - B_1^2 - \sigma A_1^2)\gamma - \beta_{1y}\Omega + \frac{1}{2}\beta_2\Omega^2 + K
\end{pmatrix}.$$

(14)

The dispersion relation that determines the MI is given by Eq. (13) for $K$ as a function of real $\Omega$. MI will occur when there are complex solutions for $K$, and the MI gain is then given by

$$g(\Omega) = |\text{Im}(K)|. \tag{15}$$

In the general case, the dispersion relation is quite involved. In the following sections, we provide explicit MI results for several special cases.

*2.3 Zero PMD and PDC: $\Gamma = 0$ and $\Delta C = 0$*

In this special case, the asymmetric CW solution given by Eqs. (2) and (3) degenerates into

$$a_{1x} = A_1 \exp(ik_1 z), \quad a_{2x} = A_2 \exp(ik_1 z), \quad a_{1y} = A_1 \exp(ik_2 z), \quad a_{2y} = A_2 \exp(ik_2 z). \tag{16}$$

with the amplitudes, propagation constants, and total power determined by the following relations:

$$A_1 A_2 = \frac{C}{\gamma(1+\sigma)}, \quad k_1 = k_2 = \frac{1}{2}\gamma(1+\sigma)P, \quad P = 2(A_1^2 + A_2^2). \tag{17}$$

The dispersion relation Eq. (13) is simplified to

$$\left\{\left[\left(K - \Omega(\beta_{1x} + \frac{\sqrt{2}}{2}C_1)\right)^2 - r_1\right]\left[\left(K - \Omega(\beta_{1x} - \frac{\sqrt{2}}{2}C_1)\right)^2 - r_2\right] - r_3\right\}$$

$$\times \left\{\left[\left(K - \Omega(\beta_{1x} + \frac{\sqrt{2}}{2}C_1)\right)^2 - r_4\right]\left[\left(K - \Omega(\beta_{1x} - \frac{\sqrt{2}}{2}C_1)\right)^2 - r_5\right] - r_6\right\} = 0, \tag{18}$$

with

$$r_1 = \frac{1}{4}\beta_2^2\Omega^4 + (\frac{1}{2}C_1^2 + \sqrt{2}\beta_2 C)\Omega^2 + 2C^2(R^2 - 1),$$

$$r_2 = \frac{1}{4}\beta_2^2\Omega^4 + (\frac{1}{2}C_1^2 - \sqrt{2}\beta_2 C)\Omega^2 + 2C^2(R^2 - 1),$$

$$r_3 = \frac{1}{2}\beta_2^2\Omega^6 C_1^2 - \left[5\beta_2^2 C^2 + C_1^4 - 4\beta_2^2 C^2 R^2\right]\Omega^4 - 8C^2(R^2 - 1)(C\beta_2 R + C_1^2)\Omega^2 + 4C^4(R^2 - 1)^2,$$

$$r_4 = \frac{1}{4}\beta_2^2\Omega^4 + \left(\frac{1}{2}C_1^2 + \sqrt{2}\beta_2 C - 2RC\beta_2 \frac{\sigma}{\sigma+1}\right)\Omega^2 + 2C^2\left(R^2 - \frac{1-\sigma}{1+\sigma}\right) - 4\sqrt{2}C^2 R\frac{\sigma}{\sigma+1},$$

$$r_5 = \frac{1}{4}\beta_2^2\Omega^4 + \left(\frac{1}{2}C_1^2 - \sqrt{2}\beta_2 C - 2RC\beta_2 \frac{\sigma}{\sigma+1}\right)\Omega^2 + 2C^2\left(R^2 - \frac{1-\sigma}{1+\sigma}\right) + 4\sqrt{2}C^2 R\frac{\sigma}{\sigma+1},$$

$$r_6 = \frac{1}{2}\beta_2^2\Omega^6 C_1^2 - \left[\beta_2^2 C^2 \frac{4+(1+\sigma)^2}{(1+\sigma)^2} - \frac{4\beta_2^2 C^2 R^2}{(\sigma+1)^2} + C_1^2(C_1^2 + 4\beta_2 CR\frac{\sigma}{\sigma+1})\right]\Omega^4$$

$$-8C^2\left(\frac{\beta_2 CR^3}{1+\sigma} - \beta_2 CR + R^2 C_1^2 \frac{1-\sigma}{1+\sigma} - \frac{C_1^2}{1+\sigma}\right)\Omega^2 + 4C^2\left[R^4 - 2R^2\frac{1+3\sigma^2}{(1+\sigma)^2} + \left(\frac{1-\sigma}{1+\sigma}\right)^2\right], \tag{19}$$

$$R = \frac{P}{P_{\min}(\sigma)}, \quad P = 2(A_1^2 + A_2^2), \tag{20}$$

$$P_{\min}(\sigma) = \frac{4C}{\gamma(1+\sigma)}, \tag{21}$$

The above asymmetric CW solution, which is obtained for zero PMD and PDC, does not reduce to that for a zero-birefringence TCF [22]. The values of the XPM coefficients are different for different fibers, which generally lead to different MI characteristics.

*2.4 Zero-birefringence TCF: $\sigma = 0$*

With $\sigma = 0$, the two polarized modes are uncoupled and Eq. (1) degenerates into two identical sets of coupled equations, with either set representing a zero-birefringence TCF. For either polarization, the corresponding dispersion relation from Eq. (13) reduces to

$$\left[\left(K - \Omega(\beta_1 + \frac{\sqrt{2}}{2}C_1)\right)^2 - r_1\right]\left[\left(K - \Omega(\beta_1 - \frac{\sqrt{2}}{2}C_1)\right)^2 - r_2\right] - r_3 = 0, \tag{22}$$

where $\beta_1$ can be either $\beta_{1x}$ or $\beta_{1y}$, and $r_1, r_2, r_3$ are given by Eq. (19) with

$$R = \frac{P}{P_{\min}(\sigma = 0)}, \quad P = A_1^2 + A_2^2, \tag{23}$$

$$P_{\min}(\sigma = 0) = \frac{2C}{\gamma}. \tag{24}$$

Note that the minimum power given by Eq. (24) is smaller by a factor of 2 than that obtained by setting $\sigma = 0$ in Eq. (21). The reason is that the input power for the zero-birefringence fiber ($\sigma = 0$), as defined by Eq. (23), is also smaller by the factor of 2 than that for the birefringent fiber, as defined by Eq. (20). Equations (22) and (24) have been derived elsewhere by directly solving the system of two coupled equations [22].

## 3. Numerical results

*3.1 Linear-birefringence TCF ($\sigma = 2/3$)*

We first address the MI in the anomalous dispersion regime. The following physical parameters are taken here: $\beta_2 = -0.02$ ps$^2$m$^{-1}$, $\gamma = 3$ (kW·m)$^{-1}$, $C_x = C_y = 200$ m$^{-1}$, and $\Gamma = C_1 = 0$. The range of the numerical values for the coupling coefficient in typical TCFs can be found in Ref. [26]. A coupling coefficient of 200 m$^{-1}$ corresponds roughly to a TCF with a core-to-core separation of 3 to 4 times of the core radius, operating at the wavelength 1.55 μm. To highlight the effects due to the XPM, both PMD and PDC are ignored.

Figure 1(a) shows the variation of the MI gain spectrum of the linear-birefringence TCF ($\sigma = 2/3$) with the normalized total input power, $P/P_{min}$. For comparison, Fig. 1(b) shows the same physical entities for the zero-birefringence TCF ($\sigma = 0$). An examination of Figs. 1(a) and 1(b) shows that the results are almost identical, regardless of the fact that the two systems have different threshold powers, i.e., $P_{min}(\sigma = 2/3) = 160$ kW and $P_{min}(\sigma = 0) = 133.3$ kW. The level of similarity is shown more clearly in Fig. 1(c), where the MI gain spectra of the two systems are juxtaposed at several normalized input powers. The results suggest that the MI gain spectrum of the linear-birefringence TCF can be obtained from that of the zero-birefringence TCF by a straightforward rescaling of the input power, i.e., the XPM effects can simply be taken into account by means of power rescaling. This is a surprising result, considering the fact that the linear-birefringence TCF is described by four coupled equations, while the zero-birefringence TCF is described by two coupled equations.

The effects of the PMD on the MI gain spectrum in the anomalous dispersion regime are shown in Fig. 2. The PMD in a TCF should be similar to that in a single-core fiber, so the PMD values used in these examples are taken as typical ones for a single-core fiber [15]. It is seen from Fig. 2 that PMD leads only to a slight decrease of the MI gain. For a realistic TCF, the value of PDC is within ±0.1% of the coupling coefficient, which can hardly affect the MI gain spectrum.

Next, we consider the MI in the normal dispersion regime, taking the following physical parameters: $\beta_2 = 0.02$ ps$^2$ m$^{-1}$, $\gamma = 6$ (kW·m)$^{-1}$, $C_x = C_y = 200$ m$^{-1}$, and $\Gamma = C_1 = 0$. Again, both PMD and PDC are set to zero to highlight the effects of XPM.

Figures 3(a) and 3(b) show the evolution of the MI gain spectra with the normalized total input power for the linear-birefringence and zero-birefringence TCFs, respectively, and Fig. 3(c) compares the MI gain spectra of the two fibers at several normalized input powers, where the threshold powers for the two fibers are $P_{min}(\sigma = 2/3) = 80$ kW and $P_{min}(\sigma = 0) = 66.7$ kW. Unlike the situation in the anomalous dispersion regime, the power-scaled MI spectra of the two fibers show obvious differences. In particular, a new MI band is generated in the linear-birefringence TCF when the input power becomes large enough, although this additional MI band is relatively insignificant, as compared to the dominant MI band.

The effects of the PMD on the MI gain spectra in the normal dispersion regime are shown in Fig. 4. The PMD tends to decrease the MI gain and can lead to splitting of the MI bands at sufficiently large values of the power, but the effects are weak over the practically relevant range of the PMD values. Similar to the situation in the anomalous dispersion regime, the PDC has negligible effects on the MI gain spectrum.

*3.2 Circular-birefringence TCF ($\sigma = 2$)*

In the anomalous dispersion regime, as in the case of the linear-birefringence TCF, the MI characteristics of the circular-birefringence TCF are almost identical to those of the zero-birefringence TCF, if the MI gain spectra of both fibers are expressed in terms of the respective normalized input powers, as shown in Fig. 5 for the same set of parameters as in Fig. 1: $\beta_2 = -0.02$ ps$^2$m$^{-1}$, $\gamma = 3$ (kW·m)$^{-1}$, $C_x = C_y = 200$ m$^{-1}$, and $\Gamma = C_1 = 0$. Note that the minimum power of the circular-birefringence TCF is $P_{min}(\sigma = 2) = 88.9$ kW, which is significantly smaller than those of the linear-birefringence TCF [$P_{min}(\sigma = 2/3) = 160$ kW] and the zero-birefringence TCF [$P_{min}(\sigma = 0) = 133.3$ kW].

The situation in the normal dispersion regime is very different. The evolution of the MI gain spectra of the circular-birefringence TCF and the zero-birefringence TCF with the respective normalized input powers are shown in Fig. 6 for $\beta_2 = 0.02$ ps$^2$m$^{-1}$, $\gamma = 6$ (kW·m)$^{-1}$, $C_x = C_y = 200$ m$^{-1}$, $\Gamma = C_1 = 0$, to allow direct comparison with the results for the linear-birefringence TCF presented in Fig. 3. The minimum power of the circular-birefringence TCF is $P_{min}(\sigma = 2) = 44.4$ kW. As shown in Fig. 6, for the circular-birefringence TCF, a new MI band appears and quickly becomes dominant with an increase of the input power. This scenario is markedly different from the situation for the zero-birefringence TCF.

In both the anomalous and normal dispersion regimes, the PMD and PDC in the circular-birefringence TCF produce only weak effects on the MI characteristics, similar to the properties exhibited in a linear-birefringence TCF.

It is worthwhile to highlight that the XPM coefficient $\sigma = 2$ case also corresponds to a different, and meaningful, physical configuration, namely, CWs of two different wavelengths being simultaneously launched into a zero-birefringence TCF. Consequently, the results obtained for the circular-birefringence TCF apply equally well to the two-wavelength case, where the PMD should be interpreted as the group-delay difference between the two wavelengths, and PDC as the wavelength-dependent coupling. A more general analysis of the MI in the two-wavelength case should also take into account the wavelength dependence of the GVD and the dispersion of the coupling coefficient.

## 4. Comparison with numerical simulations

To verify the above MI analysis, Eq. (1) was solved numerically by launching asymmetric CW states, perturbed by small-amplitude white noise, into the fiber. The equations were solved by means of the pseudospectral method in the time domain, and the fourth-order Runge-Kutta scheme with an adaptive step-size control in the space domain. The power of the added white noise was typically about 0.01% of the input CW power, and the bandwidth of the noise covered the range of [–1200 THz, 1200 THz], centered at the carrier optical frequency. In terms of the MI analysis, the maximum initial growth rate of the perturbations is expected to be in the vicinity of the dominant MI frequency, which corresponds to the maximum gain. As a result, the actual value of the dominant frequency at the onset of instability in the numerical simulations may be used to verify the MI analysis. The propagation distance is normalized by the coupling length, defined by $L_c = \pi/(2C)$, where $C_x = C_y \equiv C$ is assumed.

*4.1 The anomalous dispersion regime*

In the anomalous dispersion regime, the following fiber parameters were used, assuming a carrier wavelength of 1.5 μm: $\beta_2 = -0.02$ ps$^2$m$^{-1}$, $\gamma = 3$ (kW·m)$^{-1}$, $C_x = C_y = C = 200$ m$^{-1}$, and $\Gamma = C_1 = 0$. The minimum powers for the corresponding zero-birefringence, linear-birefringence, and circular-birefringence TCFs are, respectively, $P_{min}(\sigma = 0) = 133.3$ kW, $P_{min}(\sigma = 2/3) = 160$ kW, and $P_{min}(\sigma = 2) = 88.9$ kW.

Figures 7 – 9 display the wave-propagation dynamics for the zero-birefringence ($\sigma = 0$), linear-birefringence ($\sigma = 2/3$), and circular-birefringence ($\sigma = 2$) TCFs, respectively. In each case, the total input power normalized by the respective threshold (minimum) power is fixed at 1.2, and the power ratio between the two cores is 3.47. The period of the modulated waves at the onset of the MI can be estimated from the propagation dynamics. From the results presented in Figs. 7 – 9, the periods for the zero-birefringence, linear-birefringence, and circular-birefringence TCFs are found to be 33.5, 34.3, and 34.2 fs, respectively (all at $z = 4L_c$), which correspond to modulation frequencies 29.9, 29.2, and 29.3 THz, in good agreement with the dominant MI frequencies from the MI analysis: 29.15, 29.15 and 29.24 THz. As predicted by the MI analysis, the three fibers have almost identical MI gain profiles and should start to produce the MI at similar propagation distances.

*4.2 The normal dispersion regime*

In the normal dispersion regime, the following fiber parameters are used: $\beta_2 = 0.02$ ps$^2$m$^{-1}$, $\gamma = 6$ (kW·m)$^{-1}$, $C_x = C_y = C = 200$ m$^{-1}$, and $\Gamma = C_1 = 0$. The minimum powers for the zero-birefringence, linear-birefringence, and circular-birefringence TCFs are, respectively, $P_{min}(\sigma = 0) = 66.7$ kW, $P_{min}(\sigma = 2/3) = 80$ kW, and $P_{min}(\sigma = 2) = 44.4$ kW.

Figures 10 – 12 show the wave propagation dynamics for the single-polarization ($\sigma = 0$), linear-birefringence ($\sigma = 2/3$), and circular-birefringence ($\sigma = 2$) TCFs, respectively. In each case, the total input power normalized by the respective minimum power is fixed at 2.5, and the power ratio between the two cores is 22.96. The periods of the modulated waves for the zero-birefringence, linear-birefringence, and circular-birefringence TCFs are found to be 41.1 (at $z = 12L_c$), 43.5 (at $z = 12L_c$), and 39.9 fs (at $z = 5L_c$), which correspond, respectively, to modulation frequencies 24.3, 23.0, and 25.1 THz, in good agreement with results of the MI analysis: 24.45, 24.42, and 27.36 THz. As expected, the distance needed for the onset of the MI in the zero-birefringence and linear-birefringence TCFs are similar, which conforms to the finding from the MI analysis that the MI gain profiles are nearly identical in these two cases. On the other hand, the MI occurs at a much shorter distance in the circular-birefringence TCF, as predicted by the MI analysis, in view of the much larger growth rate.

## 5. Conclusions

In this work, we have analyzed in detail the MI characteristics of linear-birefringence and circular-birefringence TCFs for a class of asymmetric CW inputs which admit analytical solutions. We have also verified the predictions of the MI analysis by means of direct simulations of wave propagation along different fibers. The asymmetric CW state exists when the total input power exceeds a minimum value, which depends on the fiber type (zero-birefringence, linear-birefringence, or circular-birefringence). In the anomalous dispersion regime, the MI gain spectra of the three fibers are almost identical, if scaled with the respective minimum powers. The results suggest that the XPM interaction between the polarized modes of a birefringent TCF does not generate any new MI characteristics in the anomalous dispersion regime. In the normal dispersion regime, however, the power-scaled MI gain spectra corresponding to the three fibers are different and the difference is particularly significant for the circular-birefringence TCF, which has the largest XPM coefficient ($\sigma = 2$). In both regimes, the PMD and PDC show only weak effects on the MI characteristics of the TCF. Finally, it is relevant to highlight the significance of the MI in settings other than conventional optical fibers. In particular, linear couplings between parallel photonic nanowires [30–33] may display strong dispersion, which can consequently generate especially pronounced MI effects. The associated spatiotemporal solitary pulses and other related soliton phenomena [34] suggest fruitful directions of future research.

**Acknowledgements**

We appreciate a partial financial support from the Research Grants Council (Hong Kong) through contract HKU 7120/08E.

**Figures captions**

Fig. 1. (color online) Variations of the MI gain spectra with the normalized input power, $P/P_{min}$, in the anomalous dispersion regime for (a) the linear-birefringence TCF, (b) the zero-birefringence TCF, and (c) the comparison of the MI gain spectra in the linear-birefringence (dashed) and zero-birefringence (solid) TCFs at several normalized input powers. The results are obtained for the following parameters: $\beta_2 = -0.02$ ps$^2$m$^{-1}$, $\gamma = 3$ (kW·m)$^{-1}$, $C_x = C_y = 200$ m$^{-1}$, and $\Gamma = C_1 = 0$.

Fig. 2. (color online) (a) 3D and (b) 2D plots showing the variation of the MI gain spectrum with the PMD in the anomalous dispersion regime for the linear-birefringence TCF ($\sigma = 2/3$) with $\beta_2 = -0.02$ ps$^2$m$^{-1}$, $\gamma = 3$ (kW·m)$^{-1}$, $C_x = C_y = 200$ m$^{-1}$, and $C_1 = 0$. The total input power is $P = (A_1^2 + A_2^2 + B_1^2 + B_2^2) = 200$ kW.

Fig. 3. (color online) Variations of the MI gain spectra with the normalized input powers $P/P_{min}$ in the normal dispersion regime for (a) the linear-birefringence TCF, (b) the zero-birefringence TCF, and (c) the comparison of the MI gain spectra of the linear-birefringence (dashed) and zero-birefringence (solid) TCFs at several normalized input powers. The results are obtained for the following parameters: $\beta_2 = 0.02$ ps$^2$m$^{-1}$, $\gamma = 6$ (kW·m)$^{-1}$, $C_x = C_y = 200$ m$^{-1}$, and $\Gamma = C_1 = 0$.

Fig. 4. (color online) (a) 3D and (b) 2D plots showing the variation of the MI gain spectrum with the PMD in the normal dispersion regime for the linear-birefringence TCF with $\beta_2 = 0.02$ ps$^2$m$^{-1}$, $\gamma = 6$ (kW·m)$^{-1}$, $C_x = C_y = 200$ m$^{-1}$,

and $C_1 = 0$ ps m$^{-1}$. The total input power is $P = (A_1^2 + A_2^2 + B_1^2 + B_2^2) = 160$ kW.

Fig. 5. (color online) Variations of the MI gain spectra with the normalized input power $P/P_{min}$ in the anomalous dispersion regime for (a) the circular-birefringence TCF and (b) the zero-birefringence TCF, and (c) the comparison of the MI gain spectra of the circular-birefringence (dashed) and zero-birefringence (solid) TCFs at several normalized input powers. The results are obtained for the following parameters: $\beta_2 = -0.02$ ps$^2$m$^{-1}$, $\gamma = 3$ (kW·m)$^{-1}$, $C_x = C_y = 200$ m$^{-1}$, and $\Gamma = C_1 = 0$.

Fig. 6. (color online) Variations of the MI gain spectra with the normalized input power $P/P_{min}$ in the normal dispersion regime for (a) the circular-birefringence TCF and (b) the zero-birefringence TCF, and (c) the comparison of the MI gain spectra of the circular-birefringence (dashed) and zero-birefringence (solid) TCFs at several normalized input powers. The results are obtained for the following parameters: $\beta_2 = 0.02$ ps$^2$m$^{-1}$, $\gamma = 6$ (kW·m)$^{-1}$, $C_x = C_y = 200$ m$^{-1}$, and $\Gamma = C_1 = 0$.

Fig. 7. (color online) Evolution of the asymmetric CW input in a zero-birefringence TCF with $\beta_2 = -0.02$ ps$^2$m$^{-1}$, $\gamma = 3$ (kW·m)$^{-1}$, $C_x = C_y = 200$ m$^{-1}$, $C_1 = 0$, $P_{min}(\sigma = 0) = 133.3$ kW, and $P/P_{min}(\sigma = 0) = 1.2$.

Fig. 8. (color online) Evolution of an asymmetric CW input in a linear-birefringence TCF with $\beta_2 = -0.02$ ps$^2$m$^{-1}$, $\gamma = 3$ (kW·m)$^{-1}$, $C_x = C_y = 200$ m$^{-1}$, $\Gamma = C_1 = 0$, $P_{min}(\sigma = 2/3) = 160$ kW, and $P/P_{min}(\sigma = 2/3) = 1.2$.

Fig. 9. (color online) Evolution of the asymmetric CW input in a circular-birefringence TCF with $\beta_2 = -0.02$ ps$^2$m$^{-1}$, $\gamma = 3$ (kW·m)$^{-1}$, $C_x = C_y = 200$ m$^{-1}$, $\Gamma = C_1 = 0$, $P_{min}(\sigma = 2) = 88.9$ kW, and $P/P_{min}(\sigma = 2) = 1.2$.

Fig. 10. (color online) Evolution of the asymmetric CW state in a zero-birefringence TCF with $\beta_2 = 0.02$ ps$^2$m$^{-1}$, $\gamma = 6$ (kW·m)$^{-1}$, $C_x = C_y = 200$ m$^{-1}$, $C_1 = 0$, $P_{min}(\sigma = 0) = 66.7$ kW, and $P/P_{min}(\sigma = 0) = 2.5$.

Fig. 11. (color online) Evolution of the asymmetric CW state for a linear-birefringence TCF with $\beta_2 = 0.02$ ps$^2$m$^{-1}$, $\gamma = 6$ (kW·m)$^{-1}$, $C_x = C_y = 200$ m$^{-1}$, $\Gamma = C_1 = 0$, $P_{min}(\sigma = 2/3) = 80$ kW, and $P/P_{min}(\sigma = 2/3) = 2.5$.

Fig. 12. (color online) Evolution of the asymmetric CW state for a circular-birefringence TCF with $\beta_2 = 0.02$ ps$^2$m$^{-1}$, $\gamma = 6$ (kW·m)$^{-1}$, $C_x = C_y = 200$ m$^{-1}$, $\Gamma = C_1 = 0$, $P_{min}(\sigma = 2) = 44.4$ kW, and $P/P_{min}(\sigma = 2) = 2.5$.

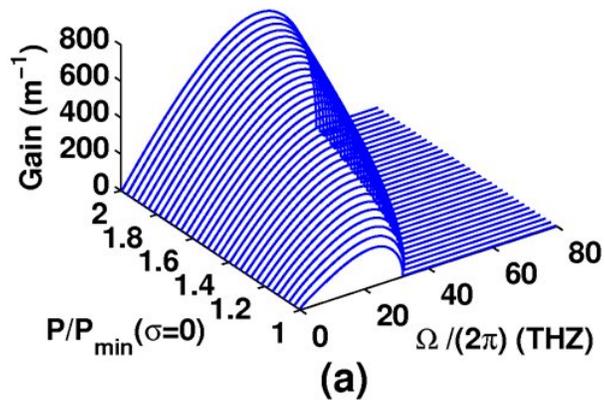

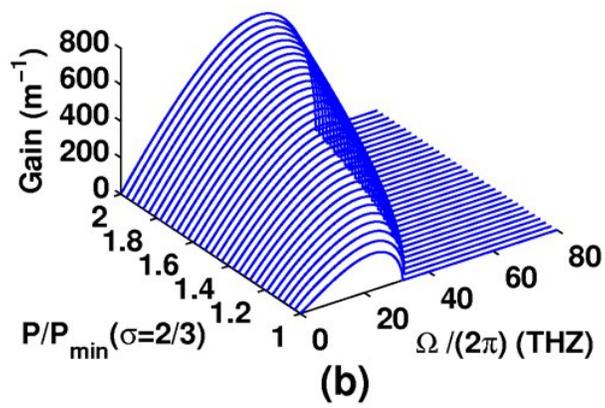

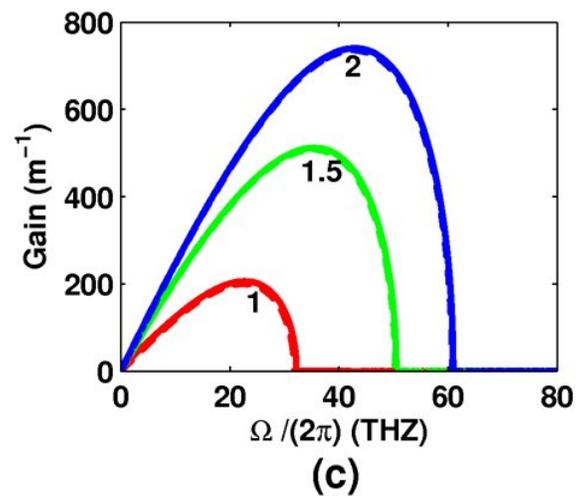

Fig. 1    J. H. Li et al

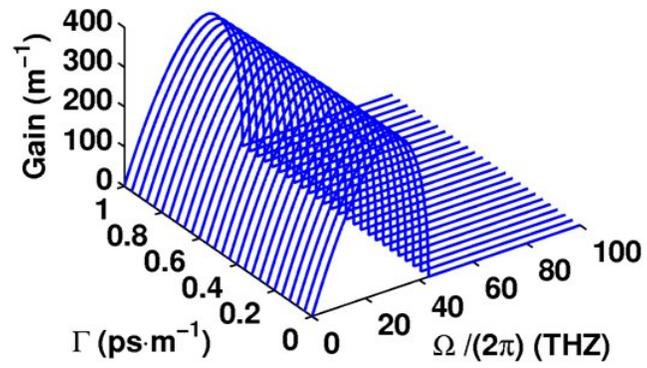

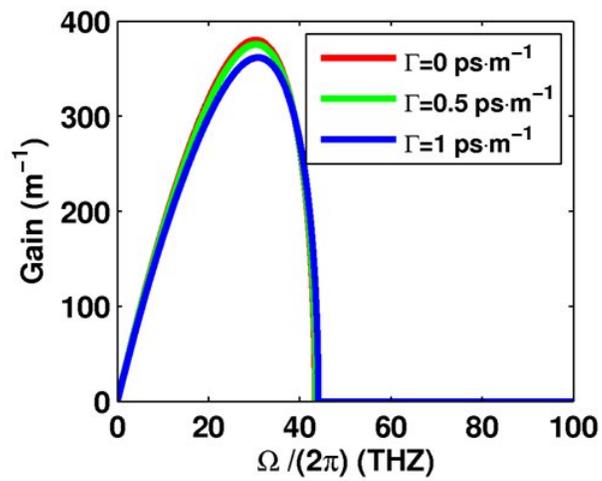

Fig. 2    J. H. Li et al

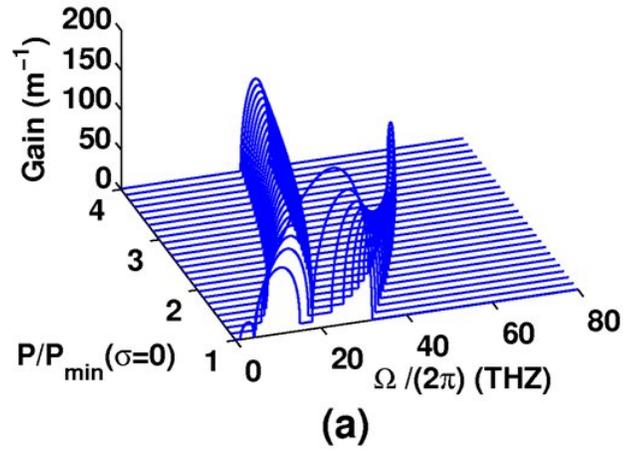

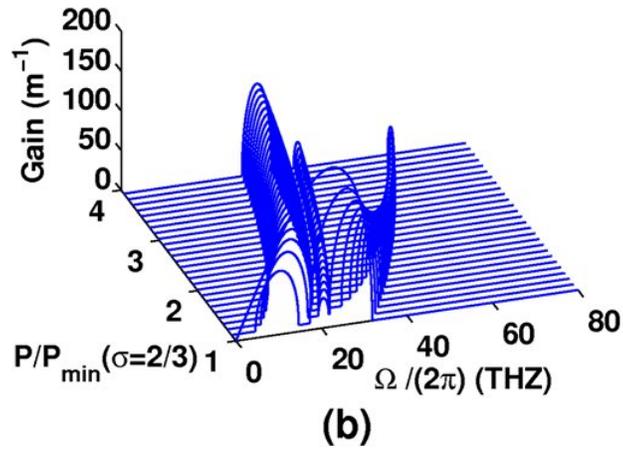

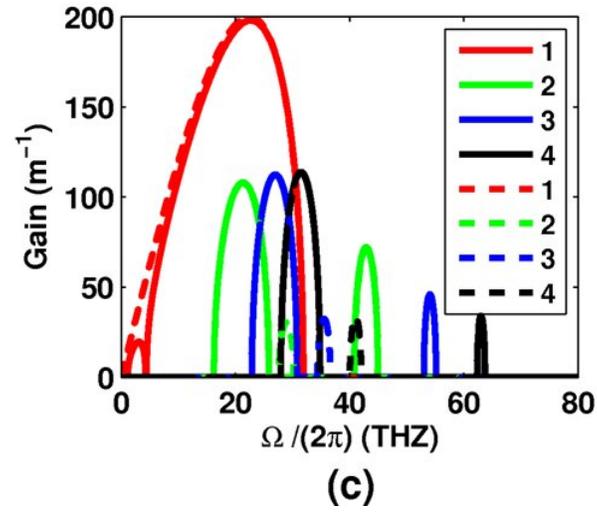

Fig. 3    J. H. Li et al

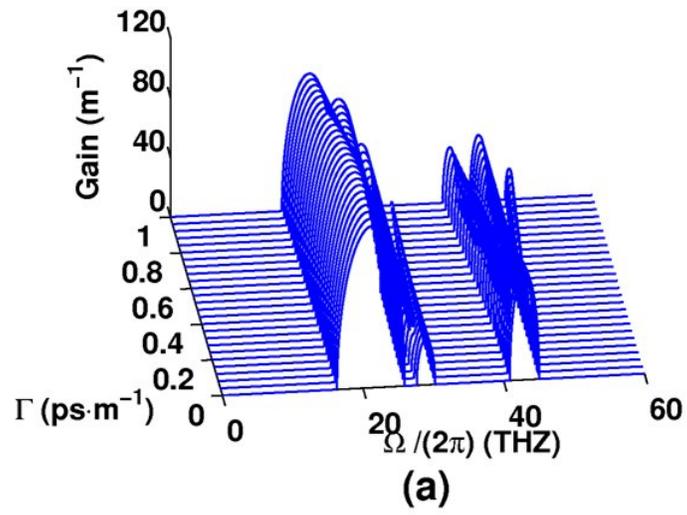

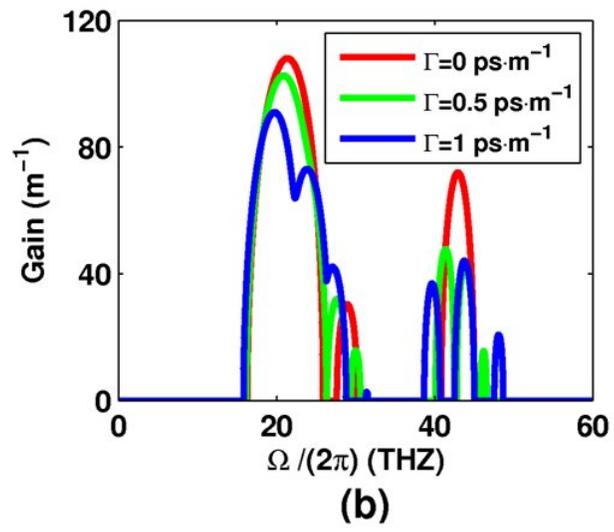

Fig. 4    J. H. Li et al

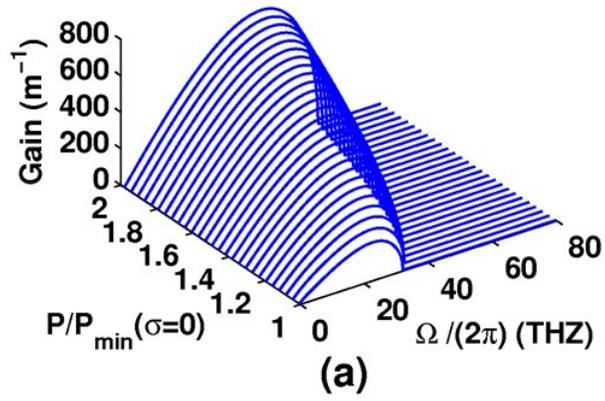

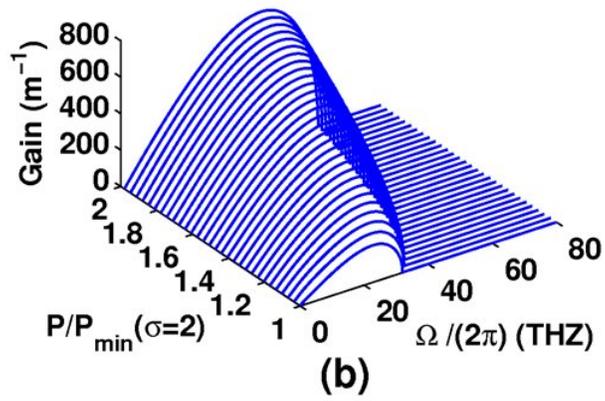

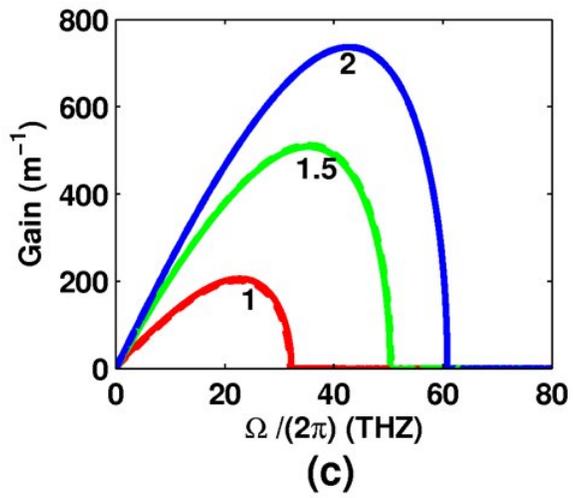

Fig. 5    J. H. Li et al

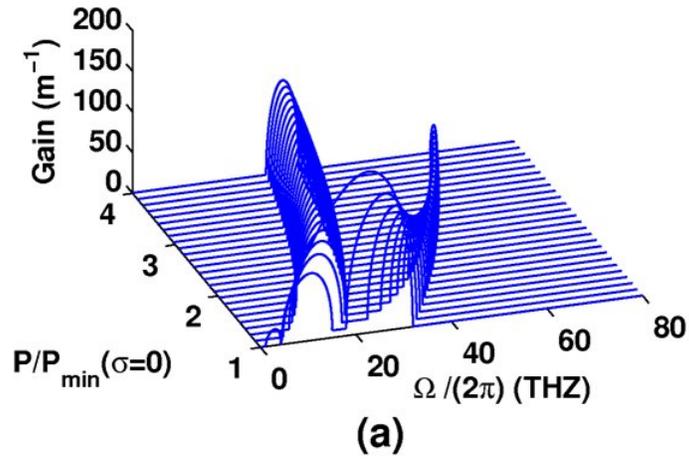

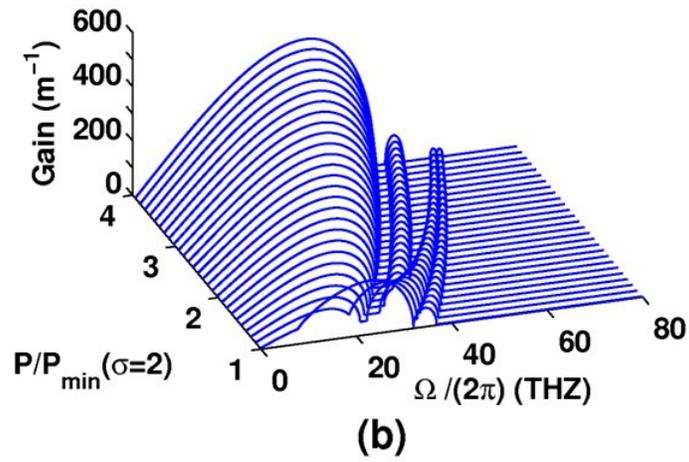

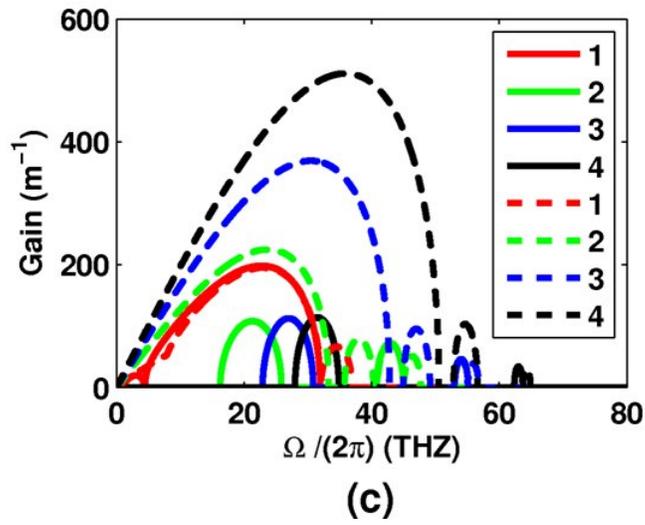

Fig. 6    J. H. Li et al

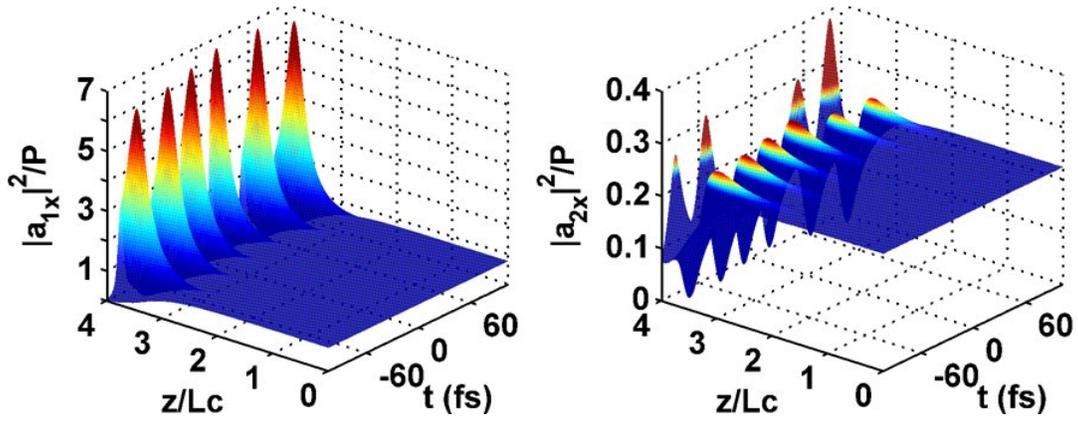

Fig. 7  J. H. Li et al

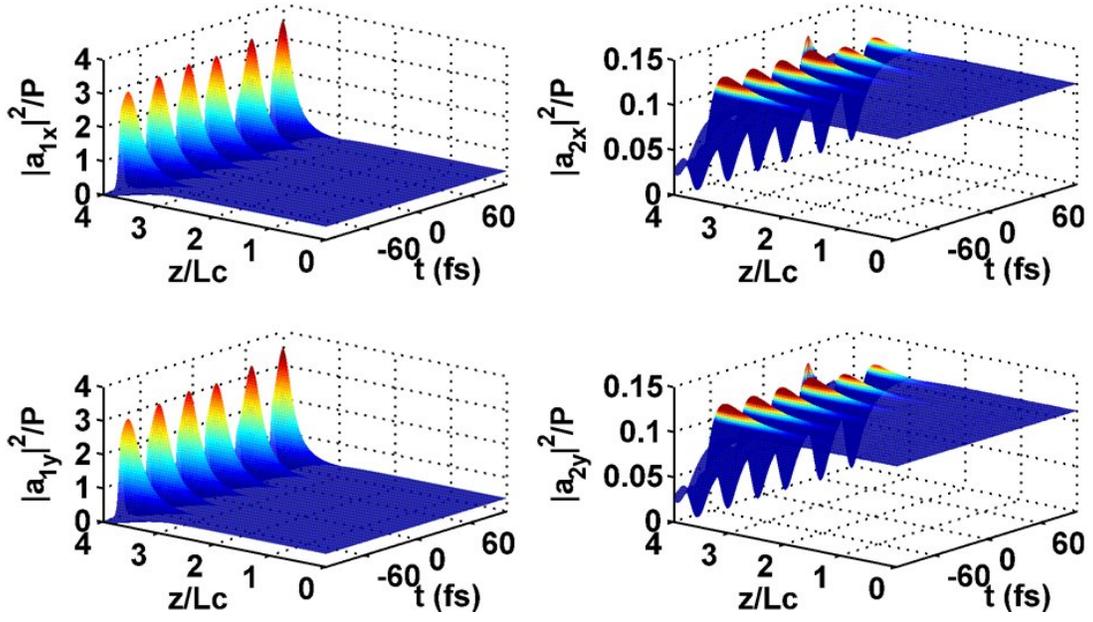

Fig. 8    J. H. Li et al

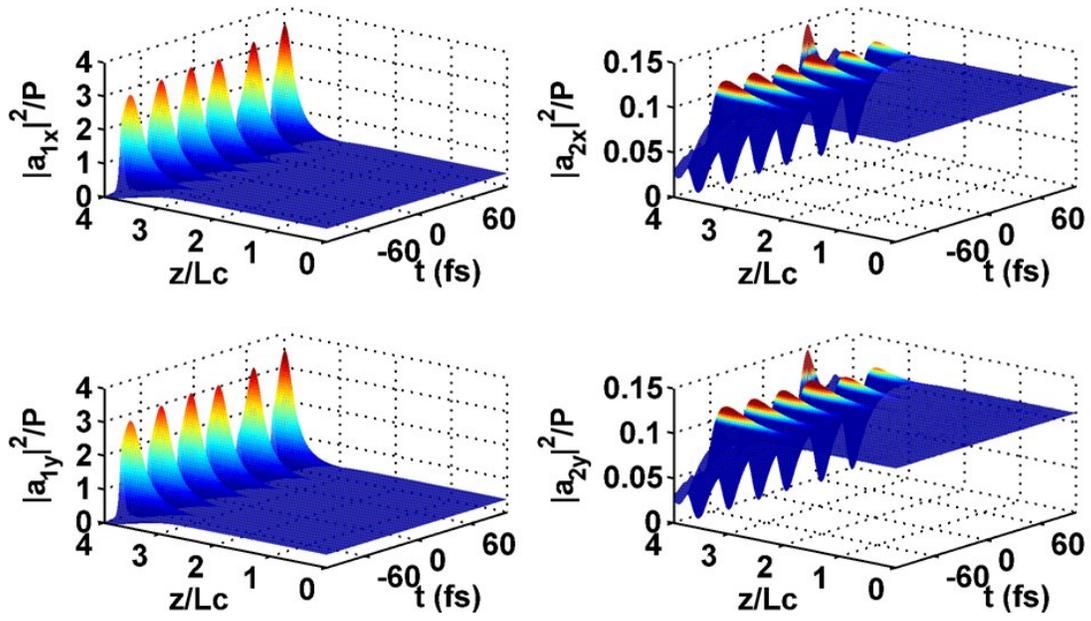

Fig. 9    J. H. Li et al

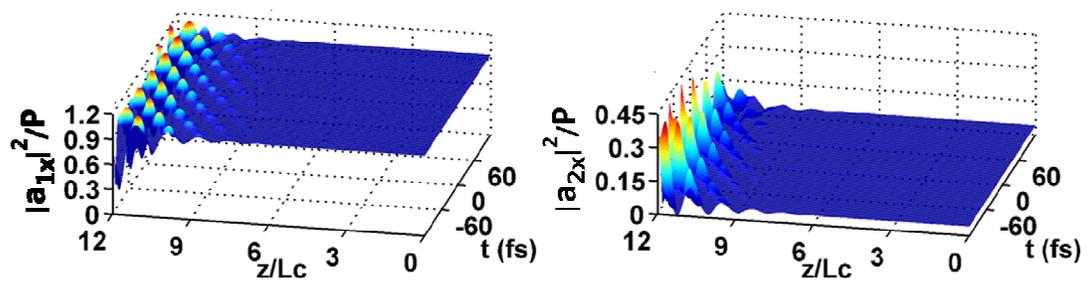

Fig. 10    J. H. Li et al

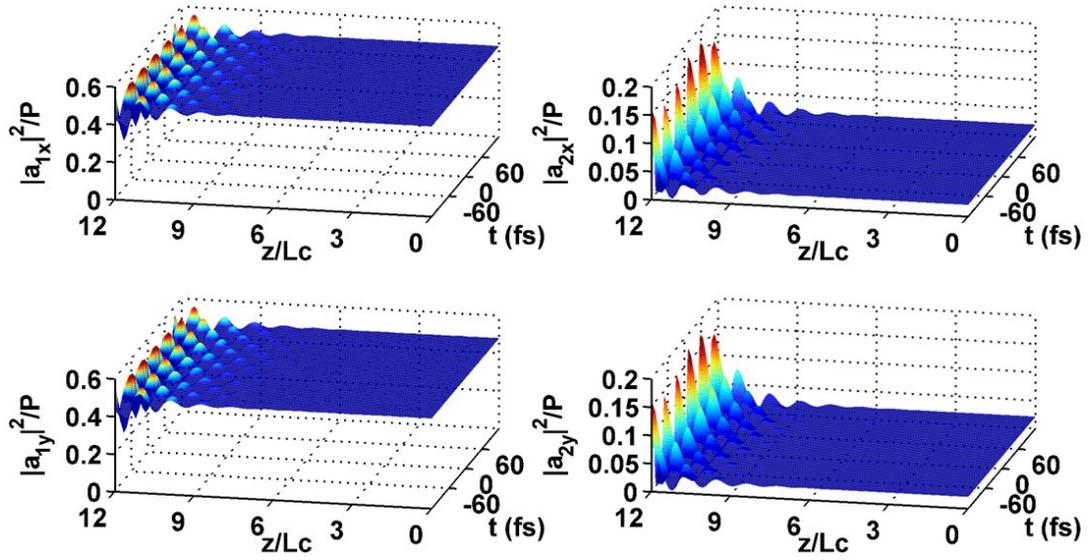

Fig. 11    J. H. Li et al

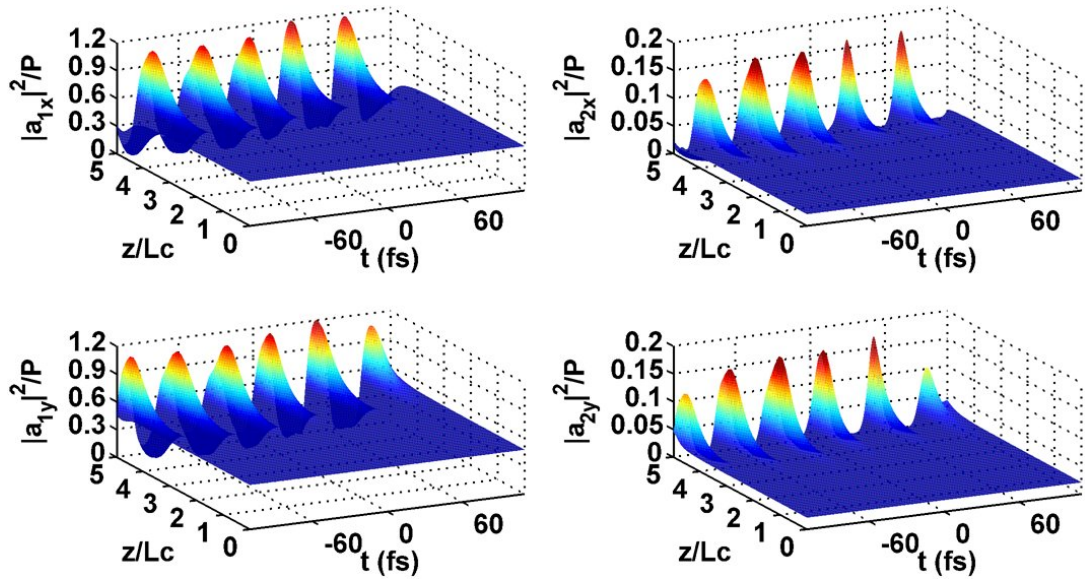

Fig. 12     J. H. Li et al